%% file: etica.tex
\documentclass{article}

\usepackage[hyphens]{url}
\usepackage{hyperref}
\usepackage{breakurl}
\usepackage{balance}
\usepackage{graphicx}
\usepackage{xspace}
\usepackage{color}
\usepackage{subfig}
\usepackage{todonotes}
\usepackage{comment}
\usepackage{array}

\author{
  Leonardo Regano, Ali Safari Khatouni, \\ Martino Trevisan, Alessio Viticchi\'e, \\
  {Politecnico di Torino, Italy}\\
  \texttt{ {firstname.lastname@polito.it} }%
}

\title{ Ethical issues of ISPs in the modern web }

\begin{document}

\maketitle

\begin{abstract}

In recent years, ethical issues in the networking field are getting more important.
In particular, there is a consistent debate about how Internet Service Providers (ISPs)
should collect and treat network measurements.

This kind of information, such as flow records,
carry interesting knowledge from multiple points of view:
research, traffic engineering and e-commerce can benefit
from measurements retrievable through inspection of network traffic.
Nevertheless, in some cases they can carry personal information
about the users exposed to monitoring,
and so generates several ethical issues.

Modern web is very different from the one we could experience few years ago;
web services converged to few protocols (i.e., HyperText Transfer Protocol (HTTP) and HTTPS)
and always bigger share of encrypted traffic.

The aim of this work is to provide an insight about
which information is still visible to ISPs in the modern web
and to what extent it carries personal information.
We show ethical issues deriving by this new situation
and provide general guidelines and best-practices to cope with the collection
of network traffic measurements.

\end{abstract}

\input{use_cases.tex}

\input{stakeholders.tex}

\input{ethical_issues}

\input{alternative_scenarios}

\bibliographystyle{abbrv}
\bibliography{etica}

\end{document}

%% file: use_cases.tex
\section{Scenario}\label{use_cases}

Passive measurement is one of the most practical ways to measure real behavior of the user and performance analysis at network or application layer. The collected data contains information about the users, services that they are using and possibly exposes content or user's credential. The latter case is related to the connections type, it is encrypted or do users use clear text credentials, etc. Basically, there are several ethical issues related to the passive measurement, e.g., which kind of data are visible at probe? What can be extract from the user's flow records? How they should store the extracted data? Who has the right to access the collected data? etc.
\newline

\begin{figure}[h]
  \begin{center}
      \includegraphics[width=0.75\textwidth]{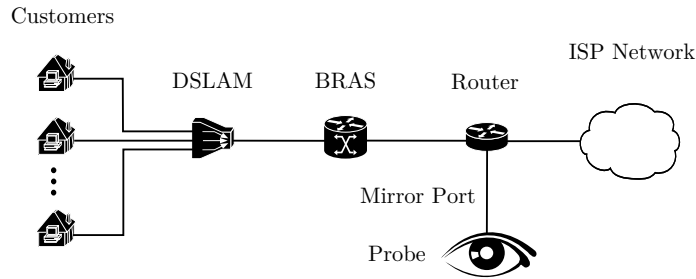}
      \caption{Typical network probe deployment in ISP.}
      \label{fig:access_network}
  \end{center}
\end{figure}

Fig.~\ref*{fig:access_network} shows the scenario in which the probe seats at ISP level, for example in a Point of Presence (PoP) where traffic of households is aggregated. All the users' connections behind the probe can be possibly captured and analysed by the ISP. Moreover, nowadays probes can filter connections or packets based on their protocol or their content. The question is, what can be seen at the ISP level, how much can be captured or stored by the probes. The storage security and accessibility of the collected data are out of scope of this work since several, generic and well-known techniques exist. However, we deal with possible ethical issues, about extracted data from users.

The probe at network layer captures traffic, such as, the IP and port of source and destination of the flows and fine grained information about the time, duration, byte transfer by each sides which provides a useful information for performance analysis for network administrators and service providers. ISPs know the personal information about the user when they apply for the service and are responsible to manage and assign clients' IP addresses. However, the mentioned data give information about where you contacted (from the IP address) and at which time and order you contacted them. The server IP address provides information about the location and possibly which service is provided by it. In particular when facing Content Delivery Networks (CDN) a single server IP address can host different services and web pages based on user geographic location and more sophisticated traffic engineering rules.
Nevertheless many techniques exist to infer the actual service used by a client in CDNs scenarios based on DNS and SSL \cite{bermudez2012dns}, allowing the ISP to an insight even on encrypted traffic going to CDN nodes which is nowadays a consistent share
\cite{trevisan_towards_2016}.

Client IP is certainly the field carrying more sensible information, helping the identification of users.
There are several techniques for IP address anonymization like \cite{crypto_pan} which is a cryptography-based sanitization tool, however the drawbacks of IP address anonymization investigated on the measurement society like \cite{anonymize_problem,anonymity}.
Transport layer related data also are distinguishable by the probe, but in this work we specifically focus on the application layer data in the modern web. 
\newline

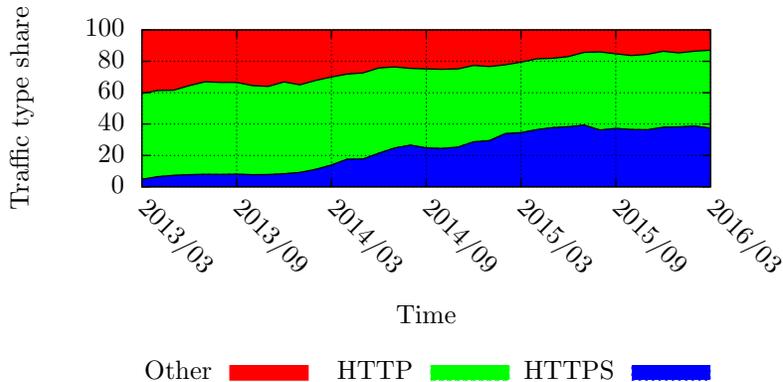
\begin{figure}[h]
  \begin{center}
      \input{figs/traffic_share/http_https.tex}
      \caption{Share of HTTP and HTTPs over last 3 years as measeured on a PoP aggregating about 10,000 households from a popular italian ISP.}
      \label{fig:http_https}
  \end{center}
\end{figure}

As depicted from the Fig.~\ref*{fig:http_https}, the portion of encrypted data on the web from 2013 to 2016 are less than 10\% and around 40\% respectively, it shows 4 times increment on 3 years. It illustrates that the service providers and users care about privacy and security to avoid readability of contents. In either HTTP and HTTPS the Domain Name System (DNS) query for resolving the name can be capture by the probe and reveal the services which users requested.
Thus, the probe can see which service the user requested \cite{bermudez2012dns}.
This data indeed represent precious information for advertisement company, service providers and market.    
Naylor et al. \cite{naylor2014cost} profoundly investigate the advantage and disadvantage of the HTTPS with respect to the HTTP. Although, HTTP traffic still have 40\% of the whole traffic share. It arises the point about how much HTTP traffic can affect users privacy? The probe sees the server IP, HTTP request and response, Uniform Resource Locator (URL) requested by client, HTTP header (cookie, referrer, ...) and  User-Agent. In particular URL can carry personal information since many servers 
encode in the URL web page parameters unveiling user actions (e.g., what he is buying on an e-commerce website).
Thus, for web services not employing encryption may lead to, consistent information exposure to the network.
Many works investigate about the ethical arguments in the modern web; in particular Zevenbergen et al.~\cite{ethical_guideline} discuss the ethical issues related to traffic measurement.

%% file: figs/traffic_share/http_https.tex
\begingroup
  \makeatletter
  \providecommand\color[2][]{%
    \GenericError{(gnuplot) \space\space\space\@spaces}{%
      Package color not loaded in conjunction with
      terminal option `colourtext'%
    }{See the gnuplot documentation for explanation.%
    }{Either use 'blacktext' in gnuplot or load the package
      color.sty in LaTeX.}%
    \renewcommand\color[2][]{}%
  }%
  \providecommand\includegraphics[2][]{%
    \GenericError{(gnuplot) \space\space\space\@spaces}{%
      Package graphicx or graphics not loaded%
    }{See the gnuplot documentation for explanation.%
    }{The gnuplot epslatex terminal needs graphicx.sty or graphics.sty.}%
    \renewcommand\includegraphics[2][]{}%
  }%
  \providecommand\rotatebox[2]{#2}%
  \@ifundefined{ifGPcolor}{%
    \newif\ifGPcolor
    \GPcolortrue
  }{}%
  \@ifundefined{ifGPblacktext}{%
    \newif\ifGPblacktext
    \GPblacktexttrue
  }{}%
  \let\gplgaddtomacro\g@addto@macro
  \gdef\gplbacktext{}%
  \gdef\gplfronttext{}%
  \makeatother
  \ifGPblacktext
    \def\colorrgb#1{}%
    \def\colorgray#1{}%
  \else
    \ifGPcolor
      \def\colorrgb#1{\color[rgb]{#1}}%
      \def\colorgray#1{\color[gray]{#1}}%
      \expandafter\def\csname LTw\endcsname{\color{white}}%
      \expandafter\def\csname LTb\endcsname{\color{black}}%
      \expandafter\def\csname LTa\endcsname{\color{black}}%
      \expandafter\def\csname LT0\endcsname{\color[rgb]{1,0,0}}%
      \expandafter\def\csname LT1\endcsname{\color[rgb]{0,1,0}}%
      \expandafter\def\csname LT2\endcsname{\color[rgb]{0,0,1}}%
      \expandafter\def\csname LT3\endcsname{\color[rgb]{1,0,1}}%
      \expandafter\def\csname LT4\endcsname{\color[rgb]{0,1,1}}%
      \expandafter\def\csname LT5\endcsname{\color[rgb]{1,1,0}}%
      \expandafter\def\csname LT6\endcsname{\color[rgb]{0,0,0}}%
      \expandafter\def\csname LT7\endcsname{\color[rgb]{1,0.3,0}}%
      \expandafter\def\csname LT8\endcsname{\color[rgb]{0.5,0.5,0.5}}%
    \else
      \def\colorrgb#1{\color{black}}%
      \def\colorgray#1{\color[gray]{#1}}%
      \expandafter\def\csname LTw\endcsname{\color{white}}%
      \expandafter\def\csname LTb\endcsname{\color{black}}%
      \expandafter\def\csname LTa\endcsname{\color{black}}%
      \expandafter\def\csname LT0\endcsname{\color{black}}%
      \expandafter\def\csname LT1\endcsname{\color{black}}%
      \expandafter\def\csname LT2\endcsname{\color{black}}%
      \expandafter\def\csname LT3\endcsname{\color{black}}%
      \expandafter\def\csname LT4\endcsname{\color{black}}%
      \expandafter\def\csname LT5\endcsname{\color{black}}%
      \expandafter\def\csname LT6\endcsname{\color{black}}%
      \expandafter\def\csname LT7\endcsname{\color{black}}%
      \expandafter\def\csname LT8\endcsname{\color{black}}%
    \fi
  \fi
  \setlength{\unitlength}{0.0500bp}%
  \begin{picture}(5760.00,3024.00)%
    \gplgaddtomacro\gplbacktext{%
      \csname LTb\endcsname%
      \put(176,2168){\rotatebox{-270}{\makebox(0,0){\strut{}Traffic type share}}}%
      \put(3220,594){\makebox(0,0){\strut{}Time}}%
    }%
    \gplgaddtomacro\gplfronttext{%
      \csname LTb\endcsname%
      \put(1608,173){\makebox(0,0)[r]{\strut{}Other}}%
      \csname LTb\endcsname%
      \put(3123,173){\makebox(0,0)[r]{\strut{}HTTP}}%
      \csname LTb\endcsname%
      \put(4638,173){\makebox(0,0)[r]{\strut{}HTTPS}}%
      \csname LTb\endcsname%
      \put(946,1577){\makebox(0,0)[r]{\strut{} 0}}%
      \csname LTb\endcsname%
      \put(946,1813){\makebox(0,0)[r]{\strut{} 20}}%
      \csname LTb\endcsname%
      \put(946,2050){\makebox(0,0)[r]{\strut{} 40}}%
      \csname LTb\endcsname%
      \put(946,2286){\makebox(0,0)[r]{\strut{} 60}}%
      \csname LTb\endcsname%
      \put(946,2523){\makebox(0,0)[r]{\strut{} 80}}%
      \csname LTb\endcsname%
      \put(946,2759){\makebox(0,0)[r]{\strut{} 100}}%
      \csname LTb\endcsname%
      \put(1078,1445){\rotatebox{-45}{\makebox(0,0)[l]{\strut{}2013/03}}}%
      \csname LTb\endcsname%
      \put(1792,1445){\rotatebox{-45}{\makebox(0,0)[l]{\strut{}2013/09}}}%
      \csname LTb\endcsname%
      \put(2506,1445){\rotatebox{-45}{\makebox(0,0)[l]{\strut{}2014/03}}}%
      \csname LTb\endcsname%
      \put(3221,1445){\rotatebox{-45}{\makebox(0,0)[l]{\strut{}2014/09}}}%
      \csname LTb\endcsname%
      \put(3935,1445){\rotatebox{-45}{\makebox(0,0)[l]{\strut{}2015/03}}}%
      \csname LTb\endcsname%
      \put(4649,1445){\rotatebox{-45}{\makebox(0,0)[l]{\strut{}2015/09}}}%
      \csname LTb\endcsname%
      \put(5363,1445){\rotatebox{-45}{\makebox(0,0)[l]{\strut{}2016/03}}}%
    }%
    \gplbacktext
    \put(0,0){\includegraphics{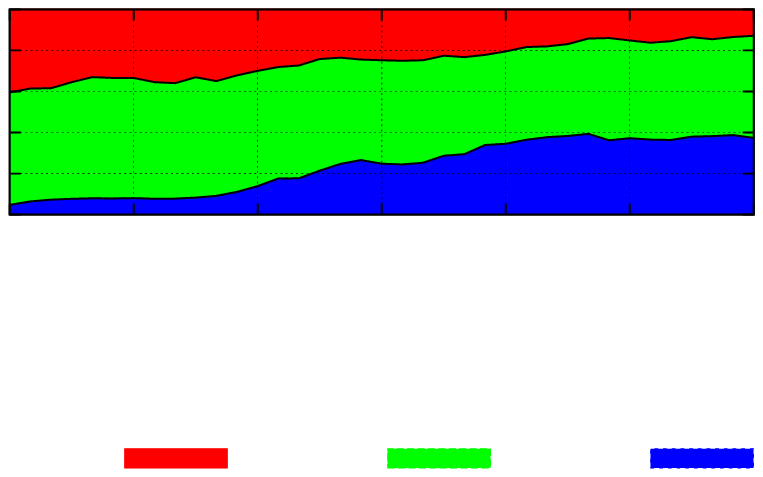}}%
    \gplfronttext
  \end{picture}%
\endgroup

%% file: stakeholders.tex
\section{Stakeholders Network}\label{stakeholders}

\begin{figure}[!t]
  \begin{center}
      \includegraphics[width=0.75\textwidth]{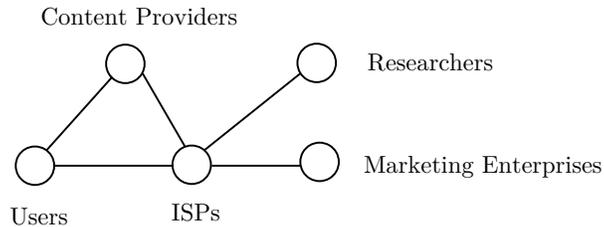}
      \caption{Involved stakeholders.}
      \label{fig:stakeholders}
  \end{center}

\end{figure}

The scenario described in the previous section depicts a very tangled and complex situation. Since many actors are involved, in this section we provide an in-depth analysis of the \emph{stakeholders} acting in this system. In Fig.\ref{fig:stakeholders} we report the stakeholders network showing involved actors and the respective interactions.

The main involved actors are without any doubt the \emph{users} of broadband Internet Access.
They are the producers of huge information used by other stakeholders, and in many case they are doing it without awareness.
Thus, ISPs' customers are not the clients but the products; personal information is a good normally sold between companies.
Users employ an \emph{Internet Service Provider} to access services across the Internet:
they expect the ISP to carries their data without looking inside the communication respecting their \emph{privacy}.
It is clear that there is a relationship of \emph{trust} between them,
and an ISP is strongly aware that the lack of trust can affect its success or bankruptcy \cite{chiou2004antecedents};
a user not trusting his provider will likely change it and that is the reason of the great effort ISPs put in loyalizing clients.
 
On the other hand, Internet Access is just a means for user to enjoy some kind of service.
Who offers a service on the Internet is called a \emph{content provider}: social networks, e-commerce portals and search-engines are content providers.
Few of them hold the majority of clients since the modern web is nowadays an oligopoly of few \emph{big players} \cite{anderson2010web}.
For example the social network field converged on two big content providers, namely \emph{Facebook} and \emph{Twitter},
while search engines are dominated by the \emph{Google} giant.
E-commerce sees few giants such as \emph{Amazon}, \emph{Ebay} and \emph{Alibaba},
whereas the first is also the owner of the biggest cloud infrastructure in the Internet.

Although the relationship between users and contents providers is evident
and well-known to the majority, ISPs and content providers are strictly connected
by often conflicting interests.
A crucial voice of cost for an ISP is represented by the traffic outgoing
from its network and, thus, it wishes to cache as much content as possible within its infrastructure.
On the other hand, content providers want to have full and exclusive visibility 
on the behavior of their users, using more and more HTTPs (encrypted) instead of HTTP,
not cacheable by its nature.
For example, \emph{Facebook} since April 2013 serves its content by means of encrypted connections:
whereas it is a significant improvement from the privacy point of view,
ISPs were certainly not happy for such decision.
Also \emph{Youtube} in January 2014 started to serve its content by means of encrypted connections,
increasing significantly the amount of traffic not cacheable within ISPs
(\emph{Youtube} traffic is more than 25\% of total).

To save part of outgoing traffic, the last trend consists in ISPs hosting within their infrastructure a Content Delivery Network (CDN) node.
A CDN is a company owning a set of cache servers across a geographical region, that Content Providers rent to serve their contents and offer their services.
Whereas an ISP can benefit from hosting a CDN node, few disadvantages are noticeable: the cache node must be powered,
and especially must be filled with content coming from outside. ISP has no control on the amount of content retrieved
and whether only its broadband clients are using that cache;
moreover, often CDNs serve the content with encryption, giving the ISP no visibility on users behavior.
Nevertheless, there are a lot of works proposing collaboration between ISPs and CDNs 
\cite{frank2013pushing}, \cite{lee2012isp}.
Again, modern fashion in content delivering is certainly less appreciated by ISPs,
but decreases the actors routing and handling with users information if compared to old-fashion cache systems owned by ISPs.

The last stakeholders involved are the consumers of network traffic information.
Several actors can benefit from information coming from the network;
for an ISP it is important to know its clients to properly engineer its infrastructure:
knowing which services are important for users allows to better configure the network to carry such traffic.
Thus, obvious questions arise: 
\begin{itemize} 
\item should an ISP \emph{know} which websites are clients visiting to improve their Quality of Experience?
\item can nowadays an ISP \emph{understand} which websites are clients visiting, the amount of mail they exchange and the videos they prefer?
\item is there the risk of incurring in severe \emph{privacy} issues (e.g., can an ISP read the content of email)?
\end{itemize} 

In the next section we focus our attention on such questions, keeping in mind that today's web is totally different from 10 years ago.
Collecting network measurement is getting harder due to encryption and Cloud/CDN infrastructure and privacy is an increasingly popular topic.
Nevertheless, what happens on the network is a rich source of information for ISPs, researchers and marketing enterprises:
many companies have as core business the collection of personal information to sell high detailed customers profile to other companies.
This kind of business concerns marketing, advertising and economy in general;
the monetary turnover of such marketplace is exponentially increasing in last years as well as users' awareness
of privacy related issues \cite{conger2013personal}.

%% file: ethical_issues.tex
\section{Ethical issues}

Given what has been depicted so far in this work, we can identify a set of
ethical issues. 
Hereafter we will discuss these issues from two main points of view:
the user one and the ISP one.

From the ISP point of view we can summarize the issue with the question: 
``is it right to access the data exchanged by a user?''.
It is not so easy to give an answer to this question: in fact, one could instinctively 
be tempted to respond: ``no, it is wrong at all'' because of the user's privacy, but there are 
some other aspects that must be considered.
In fact, it seems obvious that when a user relies on an ISP to access the global network
he should also trust the ISP and assume that his data will be shipped without any inspection.
On the other side, if an ISP could inspect user traffic, at least to apply some quality of service 
policy or just some internal routing optimizations, it would be able to improve the user network experience.
Given that, the ISP could be justified to inspect the users traffic but where is the boundary located? 
Where the ISP should stop in the inspection?
These questions comes out from the fact that the current digital world 
most advantageous activity is represented by big data analysis and users profiling \cite{chen2012business}.
Then, if an entity can access a lot of information from a large number of users it might be temped to
sell these data to the highest bidder, that is exactly the case of an ISP.
It seems useful to try to define a limit at which the inspection is deep enough for the ISP optimizations 
and not too intrusive for users privacy. 
It is not evident in which way the traffic payload inspection could be interesting for routing 
optimization purposes and then a solution could be to force ISPs to inspect only 
transport information (e.g. IP addresses, TCP/UDP ports):
nevertheless, TCP and DNS analysis are enough to assess websites a client is visiting.
The payload inspection, on the other hand, could be useful to prevent malicious intents or illegal 
communications. 
As each ISP provides the network access point for its end-users, that are typically a significant amount, 
it could be the best point to detect criminal activity, identify involved parties and prevent them. 
This fact seems to justify a deeper traffic inspection by the ISPs but, in order to protect users privacy,
the inspection activity could be delegated to a trusted third party such as police or governments 
institutions\footnote{
	In that case, it should also be discussed the trustworthiness of that entities and identified a 
	boundary between public security and a big brother effect
	}.
It is clear how the decision varies considering different aspects of the topic and how difficult 
could be to take a decision about it.

On the other hand, from the final user's point of view the issue is mainly a matter of privacy and personal 
information disclosure. 
In this case, we identified one main question: ``the user's privacy must be managed only by the user himself?''
In other words, must the end-user care about his privacy while considering all the rest of the world as untrusted
or there should be some privacy level guaranteed by the ISPs?
In this case we believe that truth lies somewhere in the middle: ISPs should implement their services in order to
not disclose users information and end-users should take care of their personal data when sent 
on the network. 
Currently, personal data protection involves data encryption that is a crucial ethical and legal point of discussion.
Recent events have underlined how encryption can cause problems in case of investigation against criminal acts, e.g.
phones used by terrorists that are totally ciphered and that the authorities cannot access (that is what they publicly said)
\cite{vice2016iphone}.
The ethical discussion about data ciphering can be very hard because it could be impossible to decide what is right among for
main possible decisions. Anyway the ethical discussion about encryption of out of the scope of this work.
We expect a great debate about this topic in the near future, where the role of network will take second place,
as major Content Providers will be involved; \emph{Whatsapp} decision to encipher all messages and hiding them from
inspection of anyone is a good example and will certainly have many consequences \cite{ieee2016whatsapp}. 

For the ISP case, data ciphering is a crucial point for an additional point: as we described in Section~\ref{use_cases} 
data anonymization could limit the ISP information about the user traffic but the sensitive data can also be accessed from 
the payload (which often is not obfuscated at all).
Then, it is clear that the user has to care of his data by ciphering payload when it is needed for privacy means but it is not
clear how to address the social security problem tied to ciphered data.
Moreover, we claim that even when having encrypted connections (i.e. HTTPs) some information is still assessable (e.g., websites history).

In conclusion, ethical issues of the ISPs in the modern web can be placed in the current digital world ethical discussion:
it is about finding a trade-off between personal privacy an public security.

%% file: alternative_scenarios.tex
\section{Alternative scenario}

As we have seen in the preceding sections, actually users have typically an high level of privacy on the payload level, but a lower one on the network level. In other words, ISPs can only know if, when, and how many times an user have visited a particular website, without having access of what activity the user have performed on it. How to conciliate the need of privacy for the user, with the technical (i.e. QoS) and commercial needs for the ISPs?

In the alternative scenario that we propose, users can voluntarily permit their ISP to access the payload of their communications. In exchange the users can enjoy globally a better service, thanks to Quality of Service (QoS): typically, we have different expectations on the waiting time to access a resource, based on the type of content we are accessing; for example, waiting some (not too much) time to access a textual web page is not a big deal, whilst having a video stream that stops constantly for buffering can be really annoying. Using QoS, ISPs can fine tune the bandwidth given to the users, in function of the accessed resources.

Users can also have economic benefits: they can share the revenues that the ISP obtained by selling their data, in terms of a reduction of their monthly fee for Internet connection. This can be also a big boost in giving access to Internet in the growing countries, and for the most disadvantaged people in first world countries: lowering the fee for accessing Internet, more people can gain access to it, thus giving more data to the ISP for them to sell, creating a virtuous cycle.

However the depicted alternative poses issues from an ethical point of view. In particular, it must be ensured that users must not be forced to ``sell'' their data to the ISPs. The risk is that ISP increase the fee to the users that do not want to give their data to them, thus making the choice practically obligatory for almost everyone. This could be avoided only by a strict control of the fees by the government regulatory agencies, thus this could be a problem in some countries, where this could be seen as an unacceptable intervention of the government in the private economy sector or even a means for political control. The ISPs can also force users to give access to their data by using the QoS, slowing Internet access to users not willing to ``sell'' their data. This behavior by ISP, although avoidable from the legal point of view (i.e. Service Level Agreement), can be hardly traceable, thus creating an ``informal'' threat to the users liberty.

Concluding, we can say that the alternative scenario proposed is feasible, but it should be deployed with great attention for the users rights, and with great scrutiny regarding the ISPs behavior.